\documentclass[aps,prl,reprint,showpacs,superscriptaddress,floatfix]{revtex4-1}
\usepackage{graphicx}
\usepackage{bm}
\begin{document}
\newcommand{\singlet}[0]
{{}^{1}S_{0}}
\newcommand{\triplet}[0]
{{}^{3}P_{2}}

\title{Dissipative Bose-Hubbard system with intrinsic two-body loss}
\author{Takafumi Tomita}
\altaffiliation{Electronic address: tomita@scphys.kyoto-u.ac.jp}
\affiliation{Department of Physics, Graduate School of Science, Kyoto University, Kyoto 606-8502, Japan}
\author{Shuta Nakajima}
\affiliation{Department of Physics, Graduate School of Science, Kyoto University, Kyoto 606-8502, Japan}
\affiliation{The Hakubi Center for Advanced Research, Kyoto University, Kyoto 606-8501, Japan}
\author{Yosuke Takasu}
\affiliation{Department of Physics, Graduate School of Science, Kyoto University, Kyoto 606-8502, Japan}
\author{Yoshiro Takahashi}
\affiliation{Department of Physics, Graduate School of Science, Kyoto University, Kyoto 606-8502, Japan}
\date{\today}

\begin{abstract}
We report an experimental study of dynamics of the metastable $^3P_2$ state of bosonic ytterbium atoms in an optical lattice. The dissipative Bose-Hubbard system with on-site two-body atom loss is realized via its intrinsic strong inelastic collision of the metastable $^3P_2$ atoms. We investigate the atom loss behavior with the unit-filling Mott insulator as the initial state and find that the atom loss is suppressed by the strong correlation between atoms. Also, as we decrease the potential depth of the lattice, we observe the growth of the phase coherence and find its suppression owing to the dissipation. 

\end{abstract}

\maketitle


In recent years, much attention has been paid to novel behaviors of cold atoms with dissipation~\cite{Daley14, Muller12, Sieberer16}. With introducing several types of dissipation, the influence of the dissipation on the quantum systems has been revealed. For example, one-body particle loss was realized by applying an electron beam~\cite{Barontini13, Labouvie15, Labouvie16} and using photon scattering process~\cite{Patil15, Luschen17}. Three-body loss was implemented by controlling the strength of three-body recombination by Feshbach resonance~\cite{Mark12}. Two-body loss process was realized by Feshbach molecules~\cite{Syassen08, Yan13}. Recently, the engineering of two-body loss in a controllable manner with the photo-association technique allows for the systematic investigation of the effect of the dissipaiton on the quantum phase transition~\cite{Tomita17}. 

Different from these rather artificial ways in introducing dissipation, a system of two-electron atoms naturally realizes the dissipative system due to the intrinsic strong inelastic collision in the metastable $^3P_2$ state~\cite{Hansen06, Traverso09, Yamaguchi08, Uetake12} and the $^3P_0$ state~\cite{Traverso09, Halder13, Franchi17, Bouganne17}. However, we have encountered the dilemma that this intrinsic strong inelastic collision also prevents previous attempts to create a Bose-Einstein condensate (BEC) and a superfluid (SF) in an optical lattice in the metastable state. Nevertheless, we have an interesting possibility of the quantum many-body physics taking advantage of the metastable state. For example, various kinds of the quantum computing platform using the metastable states for storing and controlling the quantum state are proposed~\cite{Derevianko04, Daley08, Stock08, Shibata09, Gorshkov09, Daley11, Pagano18}. With the interaction between the $^1S_0$ state and the $^3P_0$ state, two-orbital Hubbard system is investigated~\cite{Pagano15, Hofer15, Riegger18, Nakagawa18}. In the presence of the dissipation, observation of a novel quantum state is recently reported in the loss behavior of a system of the $^3P_0$ state of fermionic ytterbium isotope, consistent with the generation of a highly entangled Dicke state~\cite{Sponselee18}. 

In this Letter, we report an experimental study of dynamics of the dissipative $^3 P_2$ state of bosonic ytterbium atoms $^{174}$Yb in an optical lattice. To overcome the difficulty of making BEC in the dissipative metastable state, first we create a BEC in the ground state $^1S_0$ and form a unit-filling Mott insulator (MI) in the three-dimensional (3D) optical lattice. Then we coherently transfer the MI in the $^1S_0$ state into the $^3P_2$ state, resulting in the successful formation of the MI in the dissipative $^3P_2$ state. With this MI as an initial state, we investigate the stability of the system and find that the  atom loss is suppressed by the strong correlation. Also, this novel scheme of the initial state preparation enables us to observe the growth of the phase coherence as we decrease the lattice depth, otherwise impossible to create, and we also find that the formation of a sizable phase coherence is suppressed by the dissipation. 

The bosonic atoms in the $^3 P_2$ state in the optical lattice can be regarded as the dissipative Bose-Hubbard system described by a master equation in Lindblad form~\cite{Lindblad76}: 
\begin{equation}
\frac{d\hat{\rho}}{dt} = -\frac{i}{\hbar}[\hat{H}, \hat{\rho}] + \mathcal{L}(\hat{\rho}),
\end{equation}
where $\hat{H}$ is the Bose-Hubbard Hamiltonian 
\begin{equation}
\hat{H} = \frac{U_{ee}}{2} \sum_{j} \hat{n}_j(\hat{n}_j-1) - J \sum_{\langle j,k\rangle} (\hat{a}^{\dagger}_{j} \hat{a}_{k} + {\rm h.c.}) + \sum_{j} (\epsilon_{j}-\mu)\hat{n}_j,
\end{equation}
and $\mathcal{L}(\hat{\rho})$ represents the dissipation due to the inelastic collision between two atoms in the $^3P_2$ state
\begin{equation}
\mathcal{L}(\hat{\rho}) = \frac{\hbar\Gamma_{ee}}{4} \sum_{j} (-\hat{a}^{\dagger}_{j} \hat{a}^{\dagger}_{j} \hat{a}_{j} \hat{a}_{j}\hat{\rho} - \hat{\rho}\hat{a}^{\dagger}_{j} \hat{a}^{\dagger}_{j} \hat{a}_{j} \hat{a}_{j} + 2 \hat{a}_{j} \hat{a}_{j} \hat{\rho} \hat{a}^{\dagger}_{j} \hat{a}^{\dagger}_{j}). 
\label{eq.dissipation}
\end{equation}
$U_{ee}, J,$ and $\Gamma_{ee}$ represent the on-site interaction, the tunneling amplitude, and the inelastic collision rate, respectively. $\epsilon_j$ is the confining potential of the site $j$ and $\mu$ is the chemical potential. Index $e~(g)$ denotes ${}^3P_2 ~({}^1S_0)$ state. $\hat{a}_{j}$ is the annihilation operator of the $^3P_2$ state atoms at a site $j$ and $\hat{n}_j = \hat{a}^{\dagger}_{j} \hat{a}_{j}$. $\langle j,k \rangle$ represents nearest-neighboring pairs of lattice sites. We note that there exists the one-body loss process due to the photon scattering and the spontaneous emission, the loss rate of which is $1\sim 2$ order of magnitude smaller than the two-body loss rate.


For the full characterization of the system, it is necessary to measure the strength of the on-site interaction between the $^3P_2$ state atoms. This has never been done because of the difficulty associated with the rapid loss of atoms in the $^3P_2$ state due to the large inelastic collision. We determine the scattering length by establishing a new spectroscopic technique with double-excitation process by utilizing the inelastic loss property. 

We start with a preparation of the MI state of the $\singlet$ atoms with singly- and doubly-occupied sites at the lattice depth of $V_0 = 18~E_R$ for the $^1 S_0$ state. Here, $E_R = h^2/(2m\lambda_L^2)$ is a recoil energy, where $m$ is the mass of the $^{174}$Yb atom, $h$ is the Planck's constant and $\lambda_L = 532$~nm is the wavelength of the lattice beam. Because the polarizability of the $^3P_2$ state for the 532 nm lattice beam is different from that of the $^1S_0$ state, the lattice depth depends on the atomic state, which is taken into account in the determination of the lattice depth and the calculation of the interaction (see Supplementary material).

\begin{figure}[bt]
	\includegraphics[width=1\linewidth]{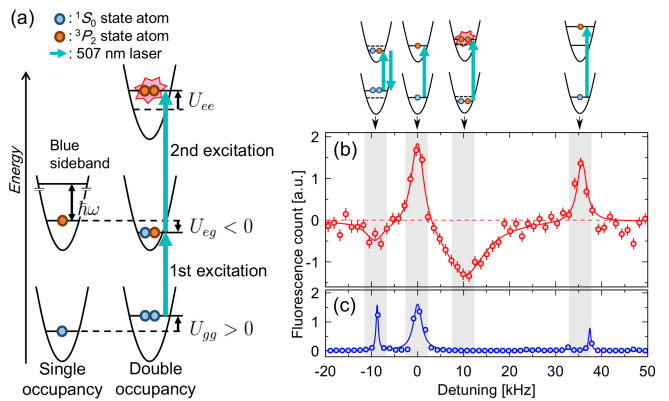}
	\caption{
		(Color online). (a) Schematic of the double-excitation spectroscopy. The on-site interaction strength manifests itself in the shift of the excitation frequency  at the doubly-occupied sites, which differs from the excitation frequency of the atoms at the singly-occupied site. (b) The spectrum of the double-excitation spectroscopy. We subtract the fluorescence count measured after the first excitation, which corresponds to the atoms excited by the first pulse. The horizontal axis represents the detuning of the second-excitation pulse frequency from the transition of the atoms in the singly-occupied sites. The resonance observed in the negative detuning corresponds the de-excitation from the $^1S_0 + {}^3P_2$ state to the $^1S_0 + {}^1S_0$ state and peak around +35 kHz represents the excitation for the blue-sideband. (c) The spectrum of the single-excitation spectroscopy for comparison. 
	}
	\label{fig_spectroscopy}
\end{figure}

We then excite a single $^1S_0$ state atom in the doubly-occupied sites into the $^3P_2$ state by adiabatic rapid passage (ARP) with a frequency-swept pulse with a 507 nm laser under a bias magnetic field of 200 mG. We perform the experiment with the atoms in the magnetic sublevel of $m_J = -2$. The atoms in the singly-occupied sites are not excited because of the well-separated resonance frequencies between the singly- and doubly-occupied sites due to the interaction (see Fig. \ref{fig_spectroscopy} (c)). Subsequently, we apply the second excitation pulse with a variable frequency. If the second pulse successfully excites a remaining $^1S_0$ state atom in the doubly-occupied sites, two $^3P_2$ state atoms occupy the same site, resulting in the strong atom loss due to the inelastic collision with the rate $\Gamma_{ee}$ (Fig. \ref{fig_spectroscopy} (a)). In the optical lattice, $\Gamma_{ee}$ is determined by the inelastic collision coefficient  $\beta_{ee}$ and the confinement of the lattice potential through the relation $\Gamma_{ee} = \beta_{ee} \int |w({\bm r})|^4 d{\bm r}$, where $w({\bm r})$ is the Wannier function of the lowest band. $\beta_{ee}$ is expected to a half of the inelastic collision coefficient with a thermal gas $\beta^{\rm thermal}_{ee} = 5.1(6) \times 10^{-11}$cm$^{3}$/s~\cite{Uetake12, Priv_Uetake15}.

Figure \ref{fig_spectroscopy} (b) shows the spectrum of the above-mentioned double-excitation spectroscopy. We observe a large dip around +10 kHz detuning from the $^1S_0 - {}^3P_2$ transition of the singly-occupied atoms, which does not have the counterpart in the spectrum of the low-intensity single pulse spectroscopy (Fig. \ref{fig_spectroscopy} (c)). We determine the interaction shifts as $(U_{ee} - U_{eg})/h = +10.7(3)$~kHz and $(U_{eg} - U_{gg})/h = -9.70(5)$~kHz. From these results and the known scattering length $a_{gg} = +104.9(1.5)a_0$~\cite{Kitagawa08}, we obtain  $a_{eg} = -201.5(1.5)a_0$ and $a_{ee} = +110(8)a_0$, where $a_0$ is the Bohr radius. This means that the on-site interaction between the $^3P_2$ atoms is repulsive and comparable to the dissipation strength: the dimensionless dissipation strength is $\hbar\Gamma_{ee}/U_{ee} = 0.94(13)$, which does not depend on the lattice depth. 

As the basic property of the dissipative quantum many-body system, we first study the stability of the unit-filling MI state in the presence of the two-body dissipation. Here, we measure the loss rate which varies as a function of the lattice depth because $J$, $U_{ee}$, and $\Gamma_{ee}$ depend on the lattice depth.

We first prepare the unit-filling MI state of the $^3P_2$ state  in almost the same manner as in the double-excitation spectroscopy, except that the lower atom number is loaded so that the doubly-occupied sites are not created. After ramping up the lattice, we excite the atoms to the $^3P_2$ state by the ARP. The remaining $^1S_0$ atoms are blasted by applying 399 nm resonant light. The atom number in the $^3P_2$ state $N(t)$ decreases as
\begin{equation}
\dot{N}(t) = - \frac{n_0 \kappa}{N(0)}N(t)^2 - \xi N(t),
\label{eq:Loss_rate}
\end{equation}
where $\kappa$ is the two-body loss rate and $\xi$ is the one-body loss rate. $n_0$ is the initial filling factor estimated by the ARP excitation efficiency, which is typically 90 \%. The one-body loss is mainly induced by the photon scattering with the $^3P_2  - {}^3S_1$ transition at 770 nm due to the 532 nm lattice beam, the rate $\gamma_{\rm sc}$ of which depends on the intensity of the lattice beam. The spontaneous emission rate is $\gamma_{\rm sp}=67(7)$~mHz \cite{Porsev99}. The one-body loss rate is given by $\xi = \gamma_{\rm sc} + \gamma_{\rm sp}$, which is calculated up to $\sim$0.3 Hz.

\begin{figure}[bt]
	\includegraphics[width=1\linewidth]{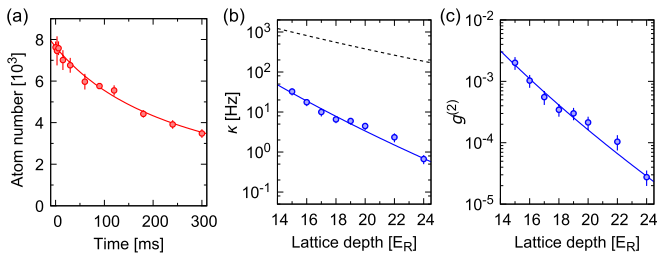}
	\caption{
		(Color online).
		(a) Time dependence of the remaining $^3P_2$ atoms at $V = 19 E_R$. The solid line shows a fit of Eq. (\ref{eq:Loss_rate}) to the experimental data.	(b)  Two-body loss rate as a function of the lattice depth. The solid line shows a fit of Eq. (\ref{eq:kappa}) to the experimental data and the dashed line is the tunneling rate for comparison. (c) Pair correlation function $g^{(2)}$ calculated from the data in the loss rate measurement. The solid line represents the theoretical calculation with $\beta_{ee}$ obtained from the data in (b). The lattice depths are adjusted for the $^3P_2$ state.  
	}
	\label{fig:kappa}
\end{figure}

Figure \ref{fig:kappa} (a) shows the typical decay of the atom number of the $\triplet$ state. By fitting Eq. (\ref{eq:Loss_rate}) to the data with the calculated one-body loss rate $\xi$, we extract the two-body loss rate $\kappa$ (see Fig. \ref{fig:kappa} (b)). The loss rate $\kappa$ is suppressed compared to the tunneling rate $6J/\hbar$ (see Fig. \ref{fig:kappa} (b)), which na\"ively characterizes the time scale of the creation of the double occupancy. In the sufficiently deep lattice, $\kappa$ can be suppressed to the order of Hz. 

This suppression is attributed to the formation of the strong correlation. When the tunneling is much smaller than the other energy scales ($J \ll \hbar\Gamma_{ee}, U_{ee}$), $\kappa$ is given by~\cite{Syassen08, Garcia09}
\begin{equation}
\kappa = \frac{16z(J/\hbar)^2}{\Gamma_{ee}} \left[1+\left(\frac{2U_{ee}}{\hbar\Gamma_{ee}}\right)^2 \right]^{-1}.
\label{eq:kappa}
\end{equation}
Here, $z = 6$ is the coordination number. We fit Eq. (\ref{eq:kappa}) to the data with the fitting parameter of $\beta_{ee}$. 
The best-fit value is $\beta_{ee} = 2.5(6) \times 10^{-11}$~cm$^3$/s, which is well agree with the half of  $\beta^{\rm thermal}_{ee} = 5.1(6) \times 10^{-11}$cm$^{3}$/s~\cite{Uetake12, Priv_Uetake15}. 

The correlation is characterized by the pair correlation function $g^{(2)} \equiv \langle \hat{n}_j(\hat{n}_j-1) \rangle/\langle \hat{n}_j \rangle ^2$, which can be estimated from the experimental result according to the relation $g^{(2)} = \kappa/\Gamma_{ee}$~\cite{Syassen08, Garcia09}. Figure \ref{fig:kappa} (c) shows that $g^{(2)}$ is much smaller than 1, which means that the creation of the double occupation is strongly suppressed. Since $g^{(2)} = 4zJ^2/[(\hbar\Gamma_{ee}/2)^2 + U_{ee}^2]$ from Eq. (\ref{eq:kappa}), the reduction of the $g^{(2)}$ is attributed to both of the on-site elastic interaction $U_{ee}$ and the inelastic loss $\hbar\Gamma_{ee}$. In our experimental parameter of $\hbar\Gamma_{ee}/U_{ee} = 0.94(13)$, the inelastic interaction contributes to the formation of the correlation in addition to the elastic interaction, although the strength of the inelastic collision does not achieve the quantum Zeno region as the previous experiments~\cite{Syassen08, Yan13, Tomita17} in which $\kappa$ decrease as the strength of the dissipation increases.

\begin{figure}[bt]
	\includegraphics[width=1\linewidth]{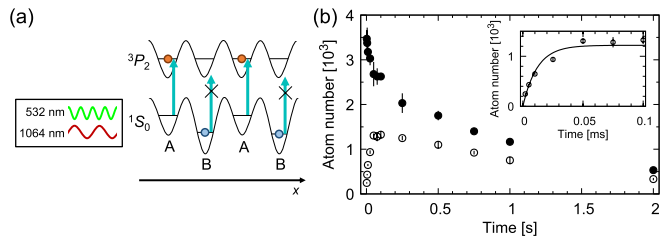}
	 \caption{
		(Color online). (a) Schematic of the selective excitation using a superlattice. The potential difference between the A and B layers created by the long lattice allows us to excite the atoms only in the A layer. (b) Time dependence of the atom number in the A layer (white circle) and B layer (black circle) at $V_0 = (19.0, 19.0, 19.9) E_R$ for the $^3P_2$ state. We note that the difference of the transfer efficiency between the A and B layers causes the remaining imbalance for longer times. The inset shows the initial 0.1 sec data of the atom number in the B layer and the fitting of the exponential function.  Note that we also observe the slow decrease of the atom number due to the two-body loss at later time that the population imbalance is already reduced. 
		}
	\label{fig:ABlayer}
\end{figure}

In order to confirm that the suppression of the doubly-occupied sites is not due to the reduction of the tunneling amplitude itself but due to the correlation effect as a result of the occupation of the atoms in the nearest neighboring site, we observe the tunneling dynamics from the initial state in which there is no atom in the nearest neighboring sites along one direction. After preparing the MI state with the $^1S_0$ atoms, we form the optical superlattice by adding the long lattice with 1064 nm laser along the $x$ axis with the relative phase between two lattice beam adjusted to make potential difference between A and B layers, which separates the excitation frequency (Fig. \ref{fig:ABlayer} (a)). We selectively excite the atoms to the $^3P_2$ state only in the A layer with ARP and blast the remaining $^1S_0$ atoms. Then we remove the additional lattice and monitor the atom number. The detection is also selectively performed with the coherent transfer to the $^1S_0$ state using ARP. We observe fast decrease of the atom number in the A layer and increase of the atom number in the B layer (Fig. \ref{fig:ABlayer} (b)), which indicates the tunneling of the atoms along the $x$ axis. From the fitting, we obtain the tunneling rate of $R=42(6)$ Hz, which is much larger than the observed $\kappa$ in the case of the unit-filling MI (see Fig. \ref{fig:kappa} (b)) and is consistent with the relaxation time scale $4J/h = 50$ Hz discussed in Ref~\cite{Trotzky12} (see Supplementary material for details).


\begin{figure}[t]
	\begin{center}
	\includegraphics[width=1\linewidth]{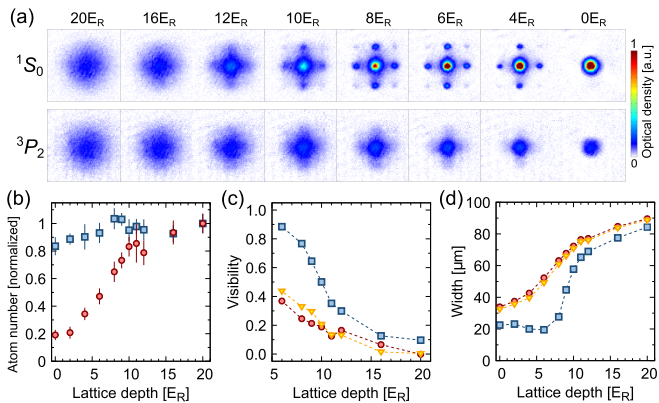}
	 \caption{
		(Color online). (a) Absorption images of the atoms. The images are taken with different final lattice depths. (b) Temporal change of the atom number during the ramp-down sequence, which is normalized by the initial atom number at the lattice depth of $V_0 = 20~E_R$. (c) Visibility of the interference peak of the images and (d) width of the density distribution. The width is the full width half maximum obtained by the Gaussian fitting. In these plots, the blue square and the red circle correspond to the data for the $^1S_0$ state and the $^3P_2$ state, respectively. The yellow triangle indicates the data for the $^3P_2$ state after eliminating the effect of the momentum kick. 
		}
	\label{fig:TOFimage}
	\end{center}
\end{figure}

We also investigate the effect of the dissipation on the quantum phase transition from the MI to the SF state. We first prepare the unit-filling MI of the $^3P_2$ state in the same manner as the preparation of the initial state of the loss rate measurement. The lattice depth is $V_0 = 20~E_R$ for the $^3P_2$ state. Then we ramp down the lattice, in which the lattice ramp-down speed is $-2~E_R$/ms. The atom number and the momentum distribution during the ramp-down dynamics are obtained from the density distribution of the time-of-flight (TOF) absorption image. After ramping down the lattice to the final lattice depth, we suddenly turn off all the trap and take the image after 6-ms expansion of the atom cloud (Fig. \ref{fig:TOFimage} (a)). The atoms in the $\triplet$ state are repumped back to the $\singlet$ state 1~ms  before taking the absorption image. For comparison, we observe the atoms in the dissipationless $\singlet$ state. We compare the two results as a function of the lattice depths because the scattering lengths $a_{ee}$ and $a_{gg}$ are almost the same within the error: $a_{ee}/a_{gg} = 1.05(7)$. The lattice depth is adjusted for each state. 

Without dissipation (the $^1S_0$ state), around $V_0 \sim 10~E_R$ we observe the transition from a MI with a broad distribution to the SF with a clear interference pattern characterizing the presence of the phase coherence, which is consistent with the theoretical value of the critical lattice depth of $V_0 = 11.29(16)~E_R$. 
On the other hand, the atom distribution of the dissipative $^3P_2$ state is modified. Although we still observe the anisotropic interference pattern in the shallow lattice region, the interference pattern is unclear. 

For the quantitative analysis, we evaluate the atom number, the visibility of the interference peaks, and the width of the atom distribution obtained by the TOF images (Fig. \ref{fig:TOFimage} (b-d)). For the $^3P_2$ system, the number of atoms starts to decrease around $V_0 = 10~E_R$. This significant atom loss reflects the start point of the melting of the MI, which creates the double occupation. The visibility of the interference peaks is defined as $v = (N_{\rm max} - N_{\rm min}) / (N_{\rm max} + N_{\rm min})$~\cite{Gerbier05}, where $N_{\rm max}$ is the sum of the atom number in the regions of first-order interference peaks, and $N_{\rm min}$ is that in the regions at the same distance from the central peak along the diagonals. In both cases, $v$ increases with the ramp-down of the lattice. This increase in the $^3P_2$ state is more moderate compared to that in the $^1S_0$ state (Fig. \ref{fig:TOFimage} (c)). In addition, the narrowing of the width of the density distribution is also more moderate in the case of the $^3P_2$ state (Fig. \ref{fig:TOFimage} (d)). These results suggest that the growth of the phase coherence is suppressed by the intrinsic on-site two-body dissipation. Similar behavior is observed in the previous experiment~\cite{Tomita17}, where the two-body loss is artificially introduced using the photo-association technique~\cite{Tomita17}. We estimate the effect of the momentum kick in the repumping process by the deconvolution analysis of the atom distribution (see the yellow triangles in Fig. 4 (c) and (d)), which shows that the effect on the TOF image is limited and does not change the whole behavior of these values qualitatively (see Supplementary material for details). 

We note that, in the case of the $^3P_2$ state, the formation of the interference pattern is still observed, which suggests the the growth of the phase coherence in the metastable state. 
Because of the strong inelastic collision, it is difficult to create the BEC in the metastable state and load it into the optical lattice. On the other hand, in our method with the slow ramp-down of the lattice, we can load the metastable atoms into the shallow optical lattice with suppressing the inelastic collision between atoms. 

In conclusion, we have realized the dissipative Bose-Hubbard system with the metastable $^3P_2$ state of $^{174}$Yb by first creating a MI state in the ground state and the subsequent coherent transfer of the atoms into the $^3P_2$ state, evading the large inelastic loss process in the state preparation. We fully characterize the system by measuring the scattering length between two $^3P_2$ atoms by developing the double-excitation method. In the 3D optical lattice, we investigate the atom loss behavior with the unit-filling MI as the initial state and find that the atom loss is suppressed by the strong correlation between atoms. Also, as we decrease the potential depth of the lattice, we observe the growth of the phase coherence and find that the formation of a sizable phase coherence is suppressed by the dissipation. 

It is expected that similar behaviors will be observed with the $^3P_0$ state of Yb~\cite{Bouganne17,Franchi17} and other two-electron atomic species~\cite{Traverso09, Hansen06, Halder13}. The strong suppression of the inelastic collision between atoms in the metastable state allows us to investigate the two-component many-body physics~\cite{Nakagawa18} and the manipulation of the $^3P_2$ atoms exploiting the magnetic dipole moment~\cite{Shibata09}, avoiding the atom loss in the practical time scale of the experiment.

\begin{acknowledgments}
We thank S. Uetake, I. Danshita and Y. Ashida for fruitfull discussions. This work was supported by the Grant-in-Aid for Scientic Research of MEXT/JSPS KAKENHI (No. 25220711, No. 17H06138, No. 18H05405, and No. 18H05228), the Impulsing Paradigm Change through Disruptive Technologies (ImPACT) program, JST CREST (No. JPMJCR1673), and Matsuo Fundation. TT acknowledges support from the JSPS (KAKENHI grant number JP16J01590). 
\end{acknowledgments}

%

%


\onecolumngrid
\clearpage
\begin{center}
\noindent\textbf{\large{Supplementary Material for \\ Dissipative Bose-Hubbard system with intrinsic two-body loss}}
\\\bigskip
Takafumi Tomita$^{1,*}$, Shuta Nakajima$^{1,2}$, Yosuke Takasu$^{1}$, and Yoshiro Takahashi$^{1}$
\\\vspace{0.1cm}
\small{$^{1}$\emph{Department of Physics, Graduate School of Science, Kyoto University, Kyoto 606-8502, Japan}}\\
\small{$^{2}$\emph{The Hakubi Center for Advanced Research, Kyoto University, Kyoto 606-8501, Japan}}\\
\end{center}
\bigskip
\bigskip
\twocolumngrid
\normalsize


\setcounter{section}{0}
\setcounter{equation}{0}
\setcounter{figure}{0}

\renewcommand{\thesection}{S\arabic{section}}
\renewcommand{\thefigure}{S\arabic{figure}}
\renewcommand{\thetable}{S\arabic{table}}
\renewcommand{\theequation}{S\arabic{equation}}


\section{Details of the spectroscopy for the measurement of the interaction strength}

\subsection{Polarizability of the $^3P_2$ state for the lattice beam}
The polarizability of the $^3P_2$ state for the 532 nm lattice beam depends on the magnetic sublevel $m_J$ and the angle between the quantization axis and the polarization of the lattice beam $\theta_{x,y,z}$. In our experiment, the polarizability of the $^3P_2$ state is $(\alpha_{e,x}, \alpha_{e,y}, \alpha_{e,z})/\alpha_g = (1.43(5), 1.43(5), 1.132(8))$. Here, $\alpha_g$ is the polarizability of the $^1S_0$ state and $(\theta_{x}, \theta_{y}, \theta_{z}) = (0^{\circ}, 0^{\circ}, 90^{\circ})$. For example, in the spectroscopy, the lattice depth of $V_0 = 18~E_R$ for the $^1S_0$ state corresponds to $(V_{0x}, V_{0y}, V_{0z}) = (25.7, 25.7, 20.4)~E_R$ for the $^3P_2$ state.

\subsection{Detail of the repumping process}

\begin{figure}[b]
	\includegraphics[width=85mm]{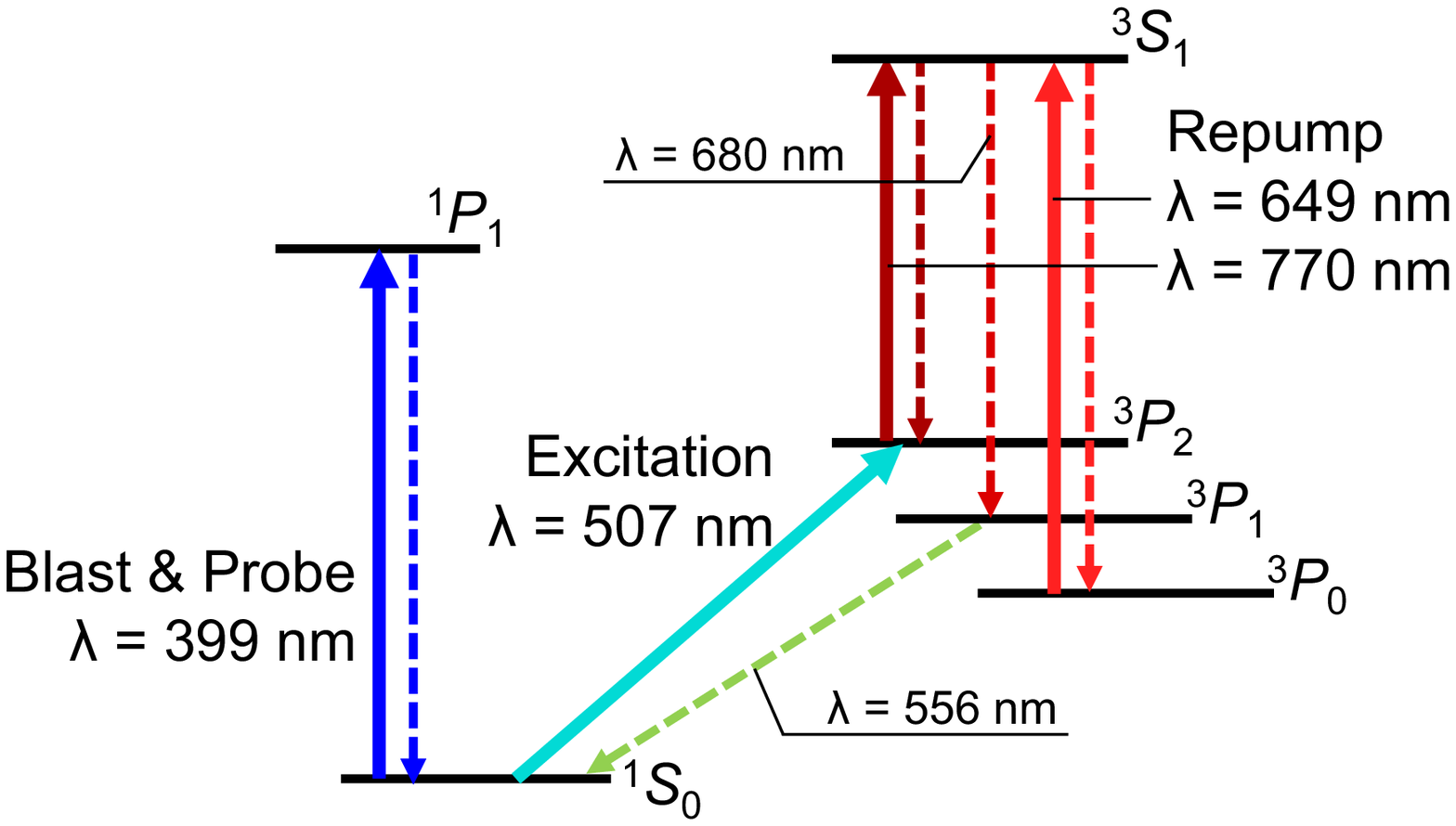}
	\caption{
		(Color online). Low-lying energy level diagram of $^{174}$Yb atoms. The solid arrows indicate the excitation transitions used in our experiment. The dashed arrows indicate the spontaneous decay lines.
		}
	\label{fig:EnergyDiagram}
\end{figure}

For the detection, atoms in the excited state $^3P_2$ are repumped back to the $^1S_0$ state using repumping lasers of 770 nm and 649 nm which are resonant to the $^3P_2 - {}^{3}S_{1}$ and $^{3}P_{0} - {}^{3}S_{1}$ transitions, respectively. The $^3P_2$ atoms absorbing a 770 nm photon is excited to the $^3S_1$ state. Then the $^3S_1$ atoms decays into the $^3P_J$ states $(J = 0,1,2)$. The atoms which decay to the $^3P_1$ state return to the $^1S_0$ state emitting 556 nm photon. The atoms which decay to the $^3P_0$ and $^3P_2$ state are again excited to the $^3S_1$ state absorbing 649 nm and 770 nm photon, respectively (see Fig. \ref{fig:EnergyDiagram}).

The repumped $^1S_0$ atoms are recaptured by a magneto-optical trap (MOT) with the $^1S_0 - {}^1P_1$ transition. The fluorescence from the MOT is detected by an electron-multiplying charge-coupled-device camera.

\section{Analysis of the tunneling behavior with the imbalanced initial state}
In general, it is difficult to calculate the behavior of the relaxation dynamics from the imbalanced initial state. In our experiments, we allow the tunneling along the $y$ and $z$ axes, thus the dynamics is rather complex. In this analysis, we simply describe the tunneling behavior as follows:
\begin{eqnarray}
\dot{N}_{A}(t) &=& -RN_{A}(t) + RN_{B}(t), \nonumber \\
\dot{N}_{B}(t) &=&  ~~RN_{A}(t) - RN_{B}(t).
\label{eq:tunneling}
\end{eqnarray}
Here, $N_{A~(B)}$ is the atom number in the A~(B) layer, and $R$ represents the tunneling rate between the A and B layers. With the initial condition that all atoms are placed in the A layer, Eq. (\ref{eq:tunneling}) yields $N_{B}(t) = [1-\exp{(-2Rt)}]N_{0}/2$, where $N_0$ is the initial atom number. We fit this function to the initial 0.1 sec data of the atom number in the B layer, as shown in the inset of Fig. 3 (b) in the main text.

\section{Estimation of the momentum kick by the repumping process in the absorption imaging}

Here we describe the effect of the momentum kick by the repumping process on the absorption imaging. Because the repumping laser is irradiated along the imaging axis, the effect of the recoil due to the photon absorption of the repumping laser is not observed. On the other hand, the expansion of the distribution of the atoms due to the recoil of the photon emission is observed because the direction of the photon emission is random and isotropic. After repumping process, the repumped $^1S_0$ state atoms expand in the accordance with the sum of the original momentum and the recoil momentum obtained by the photon emission. We estimate the width of the expansion of the atom cloud with calculating the average number of the emitted photons $N_{\rm ph}$ through the repumping process (Table \ref{table:parameter}) with the assumption that the repumping process is instantaneously finished. In the numerical calculation, we obtain the momentum distribution due to the recoil in the repumping process, which is well approximated by the gaussian function with a half width at half maximum of 1.2 $\hbar k_{L}$. Here, $k_{L} = 2\pi/\lambda_{L}$ is the wave number of the lattice beam and $\hbar k_{L}$ represents the recoil momentum of the lattice beam. 

After turning off all the trap, the $^3P_2$ atoms expands in 5 ms. Then the atoms get the recoil momenta through the repumping process and expand in 1 ms. In the numerical calculation, all atoms are repumped into the $^1S_0$ state within 10 $\mu$s, while the actual repumping time is much larger than the estimated value. We reconstruct the original atom distribution by deconvoluting the recoil momentum distribution from the atom distribution obtained from the TOF image, and estimate original visibility and width, which are shown in the yellow triangles in the Fig. 4 (c) and (d) in the main text.  

\begin{table}[t]
\caption{Information of the repump transitions~\cite{Porsev99_arXiv}. $\lambda$ is the wavelength of the transition, $\Gamma^{(s)}$ is the decay rate and $N_{\rm ph}$ is the average number of the photon emission though the repumping process. The numerical simulation is performed based on these parameters.}
\label{table:parameter}
\centering
\begin{tabular}{cccc}
\hline
\hline
                             & $^3S_1 \rightarrow {}^{3}P_0$ & $^3S_1 \rightarrow {}^3P_1$ & $^3S_1 \rightarrow {}^3P_2$ \\ 
\hline
$\lambda$ [nm] & 649 & 680 & 770 \\
$\Gamma^{(s)}$ [Hz] & 9.6 $\times 10^6$ & 2.7 $\times 10^7$ & 3.7 $\times 10^7$ \\
Branching ratio & 0.13 & 0.37 & 0.50 \\
$N_{\rm ph}$ & 1.4 & 1.0 & 0.36 \\
\hline 
\hline
\end{tabular}
\end{table}
%

%
%

%

\end{document}